# Determination of the Ignorable Boundary Condition and Standard Sample for A Novel in-situ Dynamic Mechanical Analysis Method on Soft Matter

**Longyan Wu**[1#], **Lisheng Tang**[2#], **Ran Huang**[1,2*]

[1]Academy for Engineering and Applied Technology, Fudan University, Shanghai 200433, China
[2]Department of Materials Technology and Engineering, Research Institute of Zhejiang University-Taizhou, Taizhou, Zhejiang 318000, China

#equal contribution
Correspondence to:
*huangran@fudan.edu.cn

**Abstract** An in-situ Dynamic Mechanical Analysis (DMA) method for soft matter developed by our group [Wu. et.al. 2022] encounters the problem of irregular samples, which significantly vary in shape and size in practice, therefore a standard sample "large enough" to ignore the boundary and size effects is necessary to determine the baseline of test and build the correspondence between this new method to classical mechanical tests. In this work, we use finite element analysis to approach the optimal size of a brick sample where the stress on the boundaries in three spatial directions are ignorable, and certified the results by testing a series of silicone gel samples on the in-situ DMA device. The stress-strain of tensile and compression are characterized. The material properties of gel are chosen to be close to the biological soft tissue. The size of 40mm(L)×40mm(W)×20mm(H) is determined to be the optimal result.

**1. Introduction**
   The characterizations on stress-strain properties of elastomer are very common in research and engineering for such materials, because of their characteristics such as easy deformation under force and shape recovery [1]. In material mechanics, dynamic mechanical analysis (DMA) is a classic method to obtain periodic "stress-strain" data, establish the constitutive equation, and characterize the elasticity and viscosity [2,3], especially for soft matter, e.g. rubber [4] and biological tissue [5]. There have been many reports of DMA performed on ex-vivo cadaveric specimens, and the limitations were stated [6]. Compared with the very mature standardized test methods in laboratory, on-site testing of elastomers under working conditions is still a rarely reported field, however, in real-world applications, on-site or in-situ mechanical tests are required for elastomeric materials like automobile tires that are in use, living biomaterials, or biological soft tissues that cannot be sampled and subjected to the classical methods [7,8].
   Recently our group have designed and manufactured a new set of devices and techniques to measure the in-situ stress-strain properties for soft matter, especially for the human plantar soft tissue as a biomechanical and clinical interested topic (Figure 1) [9]. However, for the in-situ material test, a significant difficulty is that the test objects encountered in the practical scenario are non-standard samples, with great variations in shape and size. Besides the results in different cases make less sense for comparability, the "apparent property" of testee characterized by in-situ method shall build a correspondence to the materials' "eigen property" conducted by classical test methods, subsequently a standard sample is necessary to be defined for the new in-situ method. This sample is required to have a regular shape and sufficient size to ignore the boundary effect, thereafter it can serve as the baseline of the material property for particular specimens. Furthermore, such a set of samples is also a must for the device calibration.

Finite element analysis (FEA), as a computational simulation, has been a widely-used tool in materials mechanics [10]. FEA offers the possibility to change the test conditions in a systematic and controlled way, such as the sample size, probing parameter, frequency and material properties. It can visually present the load distribution in soft matter, and provide internal information on the stress and strain.

Based on the above requirements, with the help of FEA, we in simulation approached the optimal size of a brick sample where the strain on the boundaries in three spatial directions are ignorable, and certified the results by testing a series of silicone gel brick samples on the in-situ DMA device. (The term "optimal" in this work means the spatial length is enough large that the boundary vertical to that direction has ignorable effect on stress propagation). The stress-strain of tensile and compression are characterized. The material properties of real sample and in FEA are chosen to be close to the biological soft tissue for our original research interest.

## 2. The methodology

### 2.1. Design of the in-situ DMA

We invented an in-situ device, as shown in Figure 1, to evaluate the multidimensional mechanical characteristics of soft matter. Based on the Dynamic Mechanical Analysis (DMA) approach, we can acquire periodic stress-strain output and analyze the viscoelasticity of soft matter by delivering dynamic stimulation to the soft matter in three types of movements (vertical, horizontal shear, and torsion). In this work for seeking the boundary condition, only vertical test (tensile and compression) is discussed.

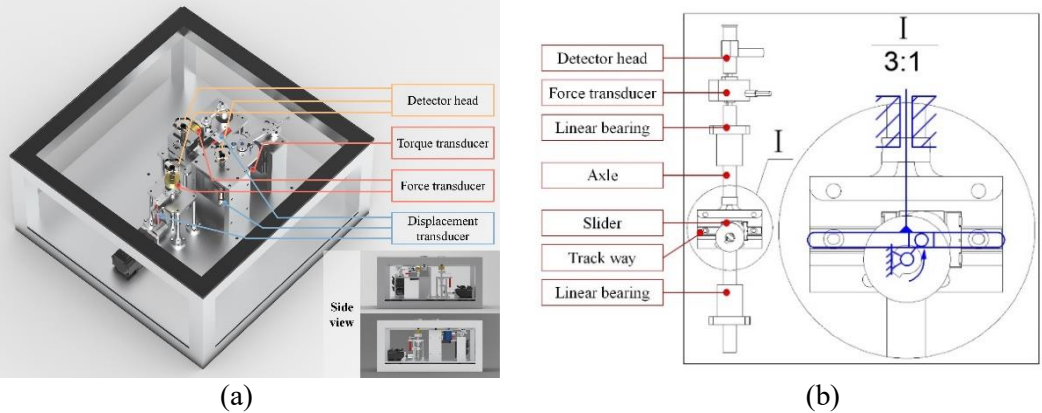

(a) (b)

Figure 1. (a)The engineering drawing and appearance of in-situ DMA device; (b) Movement sketch of the mechanical structure of the reciprocal detection with sine slider mechanism.

A detector head is attached to the testee to apply periodic sinusoidally variable stresses and deformation. The mechanical structure design of the reciprocal detector with sine slider mechanism is shown in Figure 2. Since the inspection requires non-destructive and non-invasive binding, a vacuum suction cup is employed for the detector joint. For more details of the design and device, please refer to our previous report [9].

*2.2. Samples and Test Condition*

We assume that the square brick is the appropriate shape for the standard sample, so that only the limit of the height and length on the plane need to be determined. We hope to obtain a size that when the brick is large enough to a certain extent, the stress distributed on the boundary can be small enough to be ignored. Of course, in ideal terms, the stress will propagate to infinity. According to general rules, we suppose that <5% of the central maximum stress is a negligible standard.

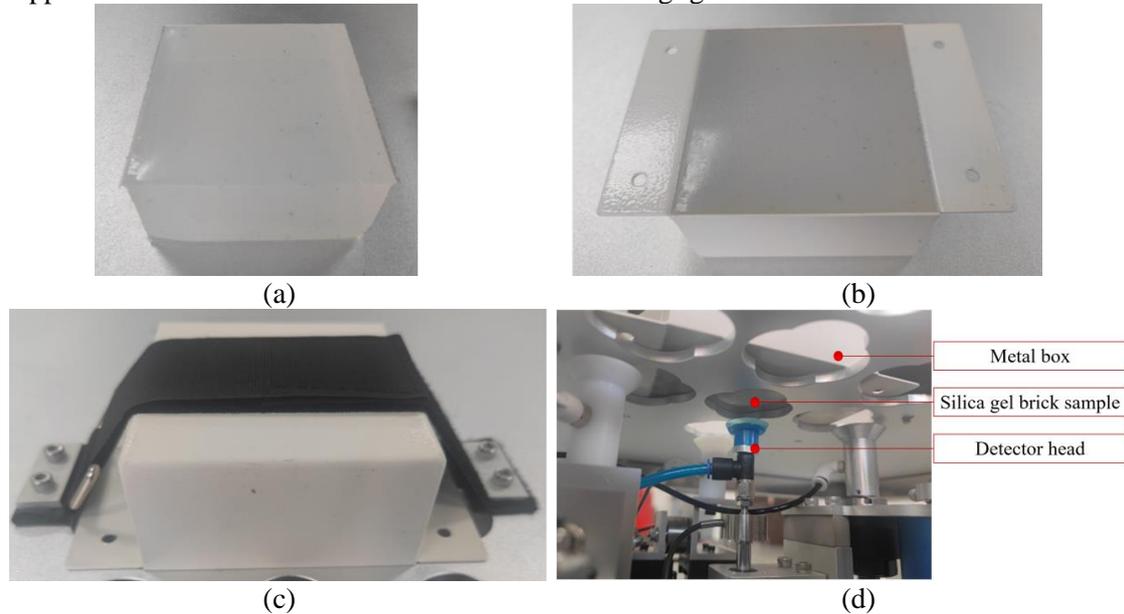

(a)　　　　　　　　　　　　　　　　(b)

(c)　　　　　　　　　　　　　　　　(d)

Figure 2. The photos of (a) silica gel brick sample; (b) the brick sample fixed in a metal box; (c) the metal box fixed on the machine and ready to test; (d) the test setup from undercover view.

It is important to note that, the "ignorable boundary condition" seeking here is a very subjective operation, because it significantly depends on the properties, especially the hardness, of the testee. It is easy to imagine that a very soft material may have an excellent stress dispersion, so that the stress applied on the its surface can propagate to a short distance, then it only requires a small size to be a "standard sample". And vice versa, hard materials have long stress propagation. Therefore, how to define and determine the standard sample, mainly depends on what kinds of soft matter are going to be tested. As stated in ref. [9], our original motivation is to study the biological soft tissue, especially the human plantar pad, so here to roughly mimic the plantar tissue, we selected the grade of silica gel with the tensile modulus of 0.518MPa, elongation at break of 363.4%, compressive modulus of 1.228MPa, hardness of 22 (shore A), and density of 1.086g/cm$^3$, to make the square brick samples. The brick is fixed in a square metal box of the same size, and the metal box is fixed on the machine so that the detector head is aligned with the center of the square for testing. Figure 2 shows a square brick of silicone gel, sample placed in the metal box, and the metal box fixed for testing.

A spoiler here tells the optimal size finally found by FEA is 40mm(L)×40mm(W)×20mm(H). Since the reality test is to certify the simulation results, we prepared five sizes of samples: 40mm×40mm×20mm, 40mm×40mm×15mm, 35mm×35mm×20mm, 50mm×50mm×30mm, 100mm×100mm×50mm. Considering the comfort of the testee and the general strength of the human soft tissue, the reciprocal displacement is set to be ±5 mm. And we have estimated the proper frequency on the human foot's soft tissues during daily activities to be 2Hz. In practice, higher frequency may cause noticeably discomfort on testee. The tests are conducted under gentle conditions with the temperature range of 23 ± 10°C (around room temperature). Silicone gel exhibits excellent thermal stability, which is also a reason for adopting it as our standard sample.

## 2.3. The FEA Simulation

The simulation test was conducted in ABAQUS (Abaqus, Inc.). The four components that make it up are the elastomer, fixed solid box, support table, and suction cup. A rigid cylinder ($D = 10$ mm) simplifies the suction cup, and the brick's material characteristics match those of the actual experiment. The simulation model is shown in Figure 3. The elastomer and the support table are fixed in space. The contact constraint is set between the elastomer and the support table as well as between the elastomer and the box. The binding constraint is set between the suction cup and the elastomer. With a sinusoidal curve of the amplitude of $\pm 5$ mm and the frequency of 2 Hz, the suction cup pulls and pushes the elastomer to perform the periodic tensile and compress test.

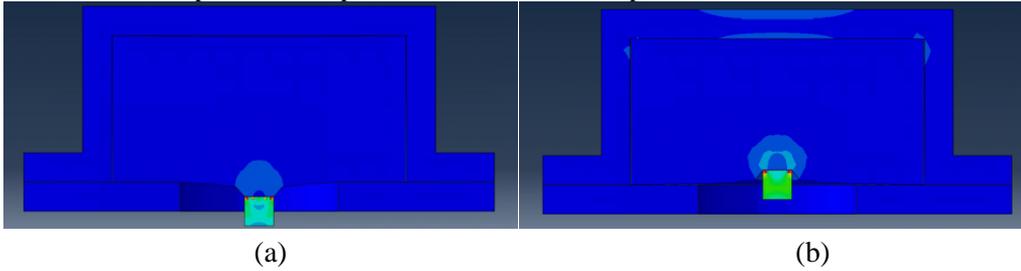

(a)          (b)

Figure 3. The simulation model: (a) tensile process and (b) compression process.

The material parameters in FEA is set to be: Density 1.086 g/cm3, Young's modulus 0.518 MPa, Poisson ratio (μ) 0.48, which is the real-measured value of the of silica gel. Fixed constraints are set at the distal end of the rectangular brick, and binding constraints are set between the probe and the elastomer. The approximate global size of the elastomer mesh is 1 mm. Hooke's law is selected to define material properties of silica gel brick sample [11]. The behaviour is described by $\sigma(\varepsilon) = E\varepsilon$, where $\sigma$ is uniaxial stress (force per unit surface), $E$ is Young's modulus, and $\varepsilon$ is strain.

## 3. Results and Discussion

### 3.1. The boundary seeking by FEA

The seeking of ignorable boundary condition is basically judged by the distribution of stresses in the cross-section. Using the legend scale in FEA software as a standard, the color block representing the smallest stress value, which is the dark blue part in Figure 5, is considered to have an approximate stress of 0. As shown in Figure 4, (a) and (b) are the cross-sectional diagrams of elastomer of length × width of 65 mm × 65 mm and height of 20 mm and 15 mm respectively. In this case, the length and width are relatively large, so it can be considered that they have almost no impact on the stress propagation to the upper direction. When the height is reduced from 20 mm to 15 mm, the stress distribution in the cross-section changes significantly, and a wide area of stress appears on the upper boundary in the 15 mm case. Only as long as the height is 20 mm, the stress on the farthest end is approximately 0. Therefore, it can be concluded that the ignorable boundary condition in height is 20 mm.

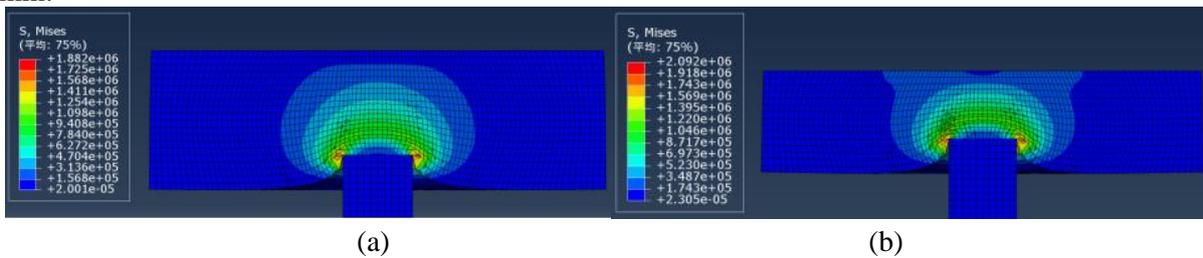

(a)          (b)

Figure 4. Cross-sectional view of the simulation with the size of the elastomer as (a) 65mm×65mm×20mm and (b) 65mm×65mm×15mm.

Similarly, the cross-sectional diagram in Figure 5 shows the simulation for the elastomer with a height of 60 mm and width of 40 mm and 35 mm. In this case, the of height is relatively large, and can be considered as "enough" on the search for the boundary of width. When the width was reduced from 40 mm to 35 mm, the stress distribution in the cross-section changed dramatically. Interestingly, in the model of 35 mm, the stress obviously propagated to the boundary, induced a significant stress expansion onto the upper direction, which makes the "safe height" of 20 mm, that we just determined above, unsecure anymore. This reminds us that the size effect in these cases is complicated and the boundary conditions in all spatial directions shall be considered simultaneously. It is concluded that the optimal boundary of length and width is 40mm×40mm.

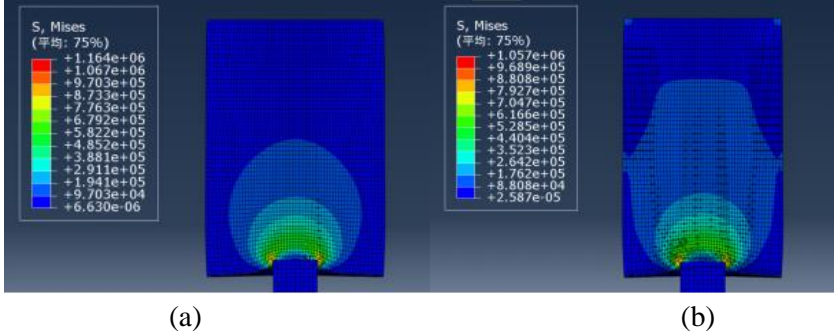

(a)   (b)

Figure 5. Cross-sectional view of the simulation results when the size of the elastomer is: (a) 40mm×40mm×60mm and (b) 35mm×35mm×60mm.

For a short conclusion, by setting 5 mm as the interval length, we found that when the length and width are relatively large (i.e., length and width have ignorable influence on the search of upper boundary), the optimal height is 20 mm; In the case of relatively large height (i.e., height has ignorable influence on the search of surrounding boundary), the optimal length and width is 40mm×40mm.

In general, the higher the analytical precision, the more computing power are needed, although the mesh density also increases analytical accuracy. The 65mm×65mm×20mm sized elastomer was re-simulated with a mesh size of 2 mm in order to increase computational efficiency, confirm the viability of the big mesh model, and study the influence of the mesh size variable on the force pattern of the elastomer in the simulation environment. Figure 6 displays the simulation results for the 2mm and 1mm mesh models, together with (a) the front view, (b) the cross-sectional view, and (c) the axial force diagram.

It can be observed that, the distributions of the stress are basically the same in both 1mm and 2mm mesh lattices, the only difference lies in the magnitude of stress values. The maximum value of stress in the model of 1mm size mesh is 1.873e+06, which is larger than the maximum of 1.252e+06 in the model of 2mm size mesh. Similarly, the minimum value of stress in 1mm model is 1.817e+02, which is much smaller than the minimum of 5.937e+02 in the other. In the cross-section diagram, the stress transition is also smoother and more stable for the small size mesh model.

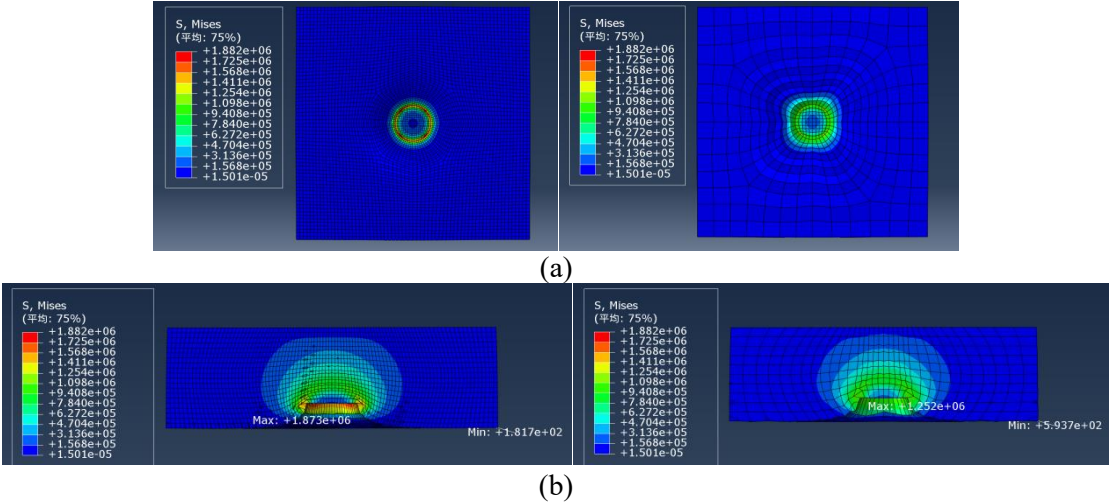

(a)

(b)

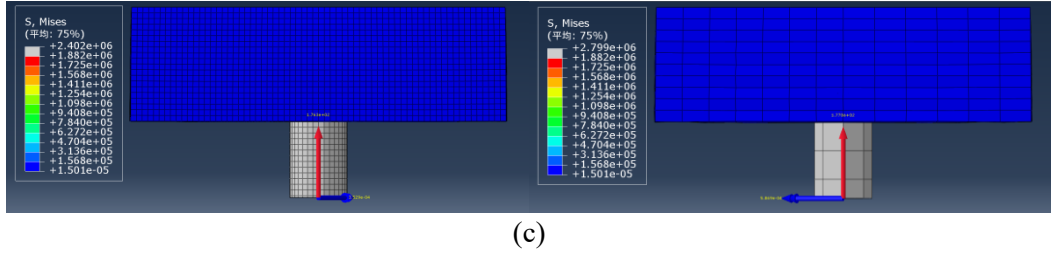

(c)

Figure 6. Comparison of simulation results for lattices of 1mm-sized and 2mm-sized meshes: (a) the front view, (b) the cross-sectional view and (c) the axial force view.

Nevertheless, the difference observed above is trivial, because the Mises stress is directly related to the size of the mesh. A force detector is set in the FEA to measure the macroscopic Newtonian force and the maximum values of axial force for 1mm-sized and 2mm-sized meshes are found to be 1.763e+02 and 1.770e+02. These two close values can convince that we can select the 2mm-sized meshes, which is capable for the simulation here to save computing power and improve efficiency, without affecting the accuracy.

*3.2. Certification with reality test*

The stress and strain behavior of a 40mm×40mm×20mm brick elastomer in both FEA and real test on silicone gel are shown in Figure 7. In real test the compression (peak) force is 37.0645±0.9318 N, the tensile (valley) force is -35.6451±0.9988 N; In FEA, the compression (peak) gives 39.7513± 1.4368 N, the tensile (valley) gives -36.7898±2.1691 N. The close results are encouraging that the FEA performs convincingly to simulate the real sample, at least in the case of silica gel brick with present material parameters.

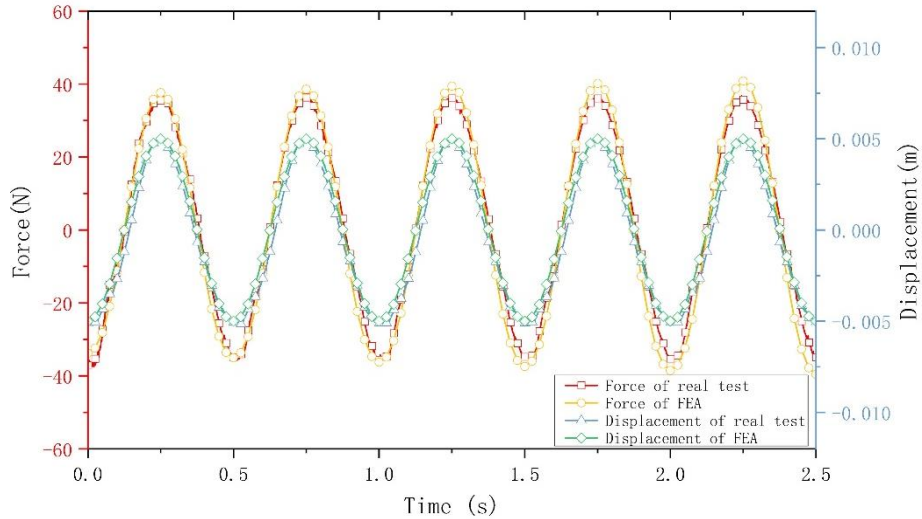

Figure 7. The stress and strain behavior of a 40mm×40mm×20mm brick elastomer in FEA and in real test on silicone brick sample.

**Table 1.** Real test results of silica gel bricks with various sizes (in mm×mm×mm)

|  | 40×40×20 (Optimal) | 40×40×15 (Less in height) | 35×35×20 (Less in width) | 50×50×30 (Larger) | 100×100×50 (Much larger) |
|---|---|---|---|---|---|
| Compression (Force, N) | 37.0645± 0.9318 | -45.6451± 1.0155 | -42.6451± 0.8984 | 36.5529± 0.8413 | 36.8742± 1.0225 |
| Tensile (Force, N) | -35.6451± 0.9988 | -43.6451± 1.0003 | -40.6451± 0.9334 | -36.3362± 0.9133 | -35.7517± 0.9885 |

As mentioned above, we have prepared five real sample of various sizes to certify the simulation. The test results are listed in Table 1, and it clearly shows the optimal samples has consistent behavior with larger and much larger samples, however the both samples with less height and width give higher stress (force) because the rigid constraint on the boundary. The results agree well with the FEA simulation and confirm that the 40mm×40mm×20mm is suitable size of standard sample for current materials.

## 4. Conclusion and Perspective

We used finite element analysis to approach the ideal size of a brick sample where the strain on the boundaries in three spatial directions are ignorable. A number of silicone brick samples were tested on the in-situ DMA device in order to validate the findings. The material specifications were chosen to closely resemble biological soft tissue. Tensile and compression stress-strain are described. The optimal size found is 40mm(L)×40mm(W)×20mm(H).

This work established a fundamental factor for the standardization of our new in-situ DMA test. With the standard sample we can build a correspondence between the in-situ result and classical methods, and furthermore we can estimate the "eigen properties" of testee, which cannot be subjected onto the classical tests e. g. living tissues, through this bridge. Here are still a number of challenges to conquer, we plan to measure a number of irregularly shaped objects with various size, to obtain a large number of data to quantize the correlation between standard sample and irregular samples with smaller size or various shapes, so as to establish a standard basis for the correction of the results when measuring the properties of irregularly shaped materials, such as the biological soft tissue.

We hope this research serve as a cornerstone to the new in-situ method and enhance its application in a wide extent.

**Acknowledgement**
This work is financially supported by the National Key Research and Development Program China (2022YFC2009500), the Medical Engineering Fund of Fudan University (yg2021-005, yg2022-008), and the RIZT Industrial Program (2022ZSS09, 2023CLG01, 2023CLG01PT).

**References**
[1] Rabanizada N, Lupberger F, Johlitz M, Lion A (2015). *Archive of Applied Mechanics*, **85** 1011-1023.
[2] Ginic-Markovic M, Choudhury N R, Dimopoulos M, Williams D R G, Matisons J (1998). *Thermochimica Acta*, **316(1)** 87-95.
[3] Guimarães C F, Gasperini L, Marques A P, Reis R L (2020). *Nature Reviews Materials*, **5(5)** 351-370.
[4] Dutta N K, Tripathy D K (1992). *Journal of applied polymer science*, **44(9)** 1635-1648.
[5] Ramadan S, Paul N, Naguib H E (2017). *Biomedical Materials*, **12(2)** 025013.
[6] Grigoriadis G, Newell N, Carpanen D, Christou A, Bull A M, Masouros S D (2017). *Journal of the mechanical behavior of biomedical materials*, **65** 398-407.
[7] Negishi T, Ito K, Kamono A, Lee T, Ogihara N (2020). *Journal of the Mechanical Behavior of Biomedical Materials*, **102** 103470.
[8] Schümann M, Odenbach S (2017). *Journal of Magnetism and Magnetic Materials*, **441** 88-92.
[9] Wu L, Zhu J, Zheng J, Geng X, He X, Tang L, Huang R, Ma X (2022). *Journal of Physics - Conference Series*, **2313** 012029.
[10] Cheung J T M, Zhang M (2005). *Archives of physical medicine and rehabilitation*, **86(2)** 353-358.
[11] Willam K (2002). *Encyclopedia of physical science and technology*, **3** 603-633.